**Giant Voltage Manipulation of MgO-based Magnetic Tunnel Junctions via Localized**

**Anisotropic Strain: a Potential Pathway to Ultra-Energy-Efficient Memory Technology**

*Zhengyang Zhao[1], Mahdi Jamali[1], Noel D'Souza[2], Delin Zhang[1], Supriyo Bandyopadhyay[3],*

*Jayasimha Atulasimha[2], and Jian-Ping Wang[1]\**

[1]Department of Electrical and Computer Engineering, University of Minnesota, Minneapolis,
MN 55455

[2]Department of Mechanical and Nuclear Engineering, Virginia Commonwealth University,
Richmond, VA 23284

[3]Department of Electrical and Computer Engineering, Virginia Commonwealth University,
Richmond, VA 23284

Strain-mediated voltage control of magnetization in piezoelectric/ferromagnetic systems is a promising mechanism to implement energy-efficient spintronic memory devices. Here, we demonstrate giant voltage manipulation of MgO magnetic tunnel junctions (MTJ) on a $Pb(Mg_{1/3}Nb_{2/3})_{0.7}Ti_{0.3}O_3$ (PMN-PT) piezoelectric substrate with (001) orientation. It is found that the magnetic easy axis, switching field, and the tunnel magnetoresistance (TMR) of the MTJ can be efficiently controlled by strain from the underlying piezoelectric layer upon the application of a gate voltage. Repeatable voltage controlled MTJ toggling between high/low-resistance states is demonstrated. More importantly, instead of relying on the intrinsic anisotropy of the piezoelectric substrate to generate the required strain, we utilize anisotropic strain produced using local gating scheme, which is scalable and amenable to practical memory applications. Additionally, the adoption of crystalline MgO-based MTJ on piezoelectric layer lends itself to high TMR in the strain-mediated MRAM devices.

\*Corresponding author. Tel: (612) 625-9509. E-mail: jpwang@umn.edu

Information storage technology is constantly challenged by an increasing demand for storage units that are small, retain information for the longest time, and dissipate miniscule amount of energy to store (write) and retrieve (read) information. Magnetic random access memory (MRAM) meets these requirements to a large extent and has been proposed as a universal storage device for computer memory.[1–3] In MRAM technology, magnetic tunneling junctions (MTJ) comprise the main storage cells. Low-energy writing of bits requires an electrically tunable mechanism to reorient the magnetization of the MTJ. However, the widely studied switching mechanisms based on utilizing current induced spin-transfer-torques (STT)[4,5] or spin-orbit-torques (SOT)[6–8] incur high energy dissipation because of the relatively large writing current density.[9,10] In recent years, several mechanisms based on using voltage to control magnetization have emerged as promising routes for ultra-low power writing of data.[11–15] Among these approaches, the strain induced control of the magnetic anisotropy in multiferroic heterostructures (a magnetostrictive layer elastically coupled with an underlying piezoelectric layer) stands out as a remarkably energy-efficient switching mechanism.[16–21] It has been widely investigated in various piezoelectric/ferromagnetic bilayer thin films[22–26] or nano-structures.[27–30] There are also several theoretical predications[31–33] that such a method will dissipate only a few atto-Joules (aJ) of energy to write data. This establishes the promise of using strain to control the resistance of an MTJ for ultra-energy-efficient memory applications.

The key for strain control of the in-plane magnetization is that the in-plane strain should be anisotropic. In most of the previous reports,[24–27,34] single crystalline piezoelectric substrates $Pb(Mg_{1/3}Nb_{2/3})_{0.7}Ti_{0.3}O_3$ (PMN-PT) with (011) orientation were utilized to generate an intrinsic anisotropic strain. However, for realistic strain-mediated MRAM, MTJ's would be grown on top of a layer of polycrystalline piezoelectric thin film deposited on a traditional Si substrate for compatibility with silicon technology.[19,32,35,36] In that case, one can no longer rely on the intrinsic anisotropy of the piezoelectric material to generate the required strain. Moreover, the integration of piezoelectric layer with MTJ stack requires a practical gating scheme to achieve high scalability, low energy dissipation and individual control, which is lacking so far.

In this paper, we demonstrate giant voltage manipulation of an MgO MTJ on PMN-PT substrate with (001) orientation. Two local gating configurations are applied to produce strong *anisotropic* strain from the *isotropic* piezoelectric layer for MTJ control. It is found that the magnetic easy axis, as well as the switching field ($H_c$) and the tunnel magnetoresistance (TMR) of the MTJ, can be efficiently controlled by strain from the underlying PMN-PT substrate generated by a gate voltage. Magnetic anisotropy can be induced either along the easy axis of the MTJ, resulting in an increase of $H_c$ by more than 4 times, or along the hard axis of the MTJ, leading to a 90° rotation of the magnetization. Moreover, we demonstrate the voltage controlled MTJ toggling between the high- and low-resistance states. Our work is fundamentally different from the previous one by Li et al.[34] Instead of relying on the intrinsic anisotropy of the piezoelectric substrate (which is not practical), our device utilizes the anisotropic strain generated via the local gating schemes and is more amenable to practical memory applications.[35,36] Moreover, the localized strain allows the control of individual MTJ's with a relative small voltage, thus enabling scalability and overcoming the substrate clamping issue.[36] The adoption of crystalline MgO as the spacer layer results in high-TMR straintronic MRAM devices. Most importantly, the side-gated MTJ prototype paves the way to realizing complete magnetization "reversal", i.e. 180° rotation of the magnetization with a voltage,[19] which is the ultimate goal of strain-based magnetization manipulation.[34,37]

A schematic of the strained-MTJ is shown in Fig. 1(a), illustrating the locations of the side and back gates for generation of the localized anisotropic strains. The structure of the MTJ stack is, from bottom to top, Ta(8)/Co$_{20}$Fe$_{60}$B$_{20}$(10)/MgO(1.8)/Co$_{20}$Fe$_{60}$B$_{20}$(4)/Ta(8) (all thicknesses are in nm) grown on a PMN-PT (001) substrate. The MTJ pillar is elliptical in shape (8μm×3μm) with its easy axis (major axis) along the *y*-direction. It is located between a pair of side gates on the top side of the substrate (as shown in the optical image in Fig. 1(b)). The back side of the substrate is contacted to form a common back gate. Using the side and back gates, one can apply an electric field (E-field) across the PMN-PT substrate to generate a broad range of strain profile. The separation between the two side gates is 40 μm to ensure their electrical isolation from the MTJ.

The piezoelectric behavior of a bare PMN-PT (001) substrate is shown in Fig. 1(c), where the in-plane strain is plotted as a function of the out-of-plane electric field, measured with a general purpose 120 Ω Constantan linear foil strain gauge (EA-06-062ED-120, Vishay Precision Group, Micro-Measurements). The strain curve under bipolar E-field poling from -8 kV/cm to +8 kV/cm (solid line) exhibits typical butterfly-like behavior, and the curve under E-field with a smaller range (dashed line) exhibits almost linear behavior with a very small hysteresis.[38] The magnetic hysteresis (M-H) loop of patterned MTJ films is obtained using vibrating sample magnetometry (VSM), as shown in Fig. 1(d), indicating that the thicker CoFeB layer is magnetically harder (with a larger coercivity), while the thinner layer is softer (with a smaller coercivity). The magnetoresistance (MR) loop of the MTJ device is also presented in Fig. 1(d), with zero gate voltage applied and the magnetic field swept along the major axis (*y*-axis). A post-annealing process at 250 °C was performed for one hour to increase the TMR ratio of the MTJ. Since neither of the CoFeB layers in the MTJ is pinned by an anti-ferromagnet, the magnetic anisotropy of both layers can be affected by the strain. We can assume the strains exerted on the soft layer and that exerted on the hard layer are very close to each other, since strain relaxation between the layers is negligible in our devices. This is confirmed by making a second sample with the positions of the hard layer and the soft layer interchanged (so that the soft layer is closer to the piezoelectric substrate), the results of which is shown in the supplementary material.[39]

In this study, we present our experimental results for two different gating scenarios: a gate voltage $V_g$ is applied either between the back gate and the bottom electrode of the MTJ (Configuration I in Fig. 2), or between the back gate and a pair of side gates (Configuration II Fig. 3). In both cases, an anisotropic strain is produced, which is highly localized in the MTJ region as illustrated in Figs. 2(b) and 3(b). But the direction of the strain profile is opposite in the two gating scenarios. Note that in our experiments positive $V_g$ corresponds to the E-field being parallel to the piezoelectric polarization (poling) direction and negative $V_g$ corresponds to the E-field being anti-parallel to the poling direction.

First we study the gating effect on MTJ in Configuration I (Fig. 2(a)). With the magnetic field swept along the $y$-axis, the MR loops under three different gate voltages $V_g = -150\,\text{V}$, $0$ and $+150\,\text{V}$ are presented in Fig. 2(c) (see supplementary material[39] for more data). At $V_g = 0$, a normal MR loop similar to that in Fig. 1(d) is obtained with sharp transitions between high- and low-resistance states. However, when a negative gate voltage $V_g = -150\,\text{V}$ is applied, the sharp transitions in the MR loop change to gradual slopes, indicating that the easy axes of both magnetic layers have rotated towards the transverse direction ($x$-direction). On the other hand, when $V_g$ is positive, the switching field increases significantly upon increasing the gate voltage suggesting enhancement of the magnetic anisotropy along the major axis ($y$-axis). The variation of the switching field of the hard CoFeB layer versus gate voltage is plotted in Fig. 2(d). It can be seen

$H_c$ increases almost linearly and becomes more than 4-fold larger when $V_g$ is increased from 0 to

+150V. In order to have a better understanding of how the magnetization of the ferromagnetic

layers are affected by the gate voltage in our devices, 3D piezoelectric finite element simulations

were performed using the COMSOL Multiphysics package. The simulated piezoelectric strain

mapping on the top surface of the substrate is presented in Fig. 2(b) with $V_g = +50\,\text{V}$. A positive

E-field applied in the out-of-plane direction produces an out-of-plane expansion ($d_{33}$) and in-plane contraction ($d_{31}$) in the substrate. Therefore, upon application of a positive $V_g$, an in-plane bi-axial strain is generated in the region beneath the stripe-shaped electrode where the voltage is applied, and the strain is compressive (negative) in both $x$- and $y$-directions. Since the electrode is long in the $y$-direction and narrow in the $x$-direction, the strain component $\varepsilon_{xx}$ along the $x$-direction is dominant, resulting in an anisotropic strain on the MTJ. We define the in-plane anisotropic strain as $\varepsilon_{xx} - \varepsilon_{yy}$. From the simulation, a strain of $\varepsilon_{xx} - \varepsilon_{yy} = -274\,\text{ppm}$ is produced on the MTJ at

$V_g = +50\,\text{V}$. Such an anisotropic strain compresses the MTJ along the $x$-direction. The strain

induced magnetic anisotropy can be expressed as $K_{me} = \frac{3}{2}\lambda\sigma$ where $\lambda$ represents the magnetostriction coefficient and $\sigma = (\varepsilon_{xx} - \varepsilon_{yy})Y$ represents the stress with $Y$ being the Young's modulus.[36,40] Considering $\lambda > 0$ for CoFeB,[41] the negative $\varepsilon_{xx} - \varepsilon_{yy}$ increases the magnetic anisotropy along the $y$-direction. As a result, the MR loops in Fig. 2(c) are significantly broadened with positive $V_g$. On the other hand, with negative $V_g$, $\varepsilon_{xx}$ is dominant over $\varepsilon_{yy}$ with a positive value (tensile), i.e. $\varepsilon_{xx} - \varepsilon_{yy} > 0$. In this case, magnetic anisotropy is induced along the $x$-direction

and the easy axis of the MTJ rotates by 90°, indicated by the slanted MR loop in Fig. 2(c) at

$V_g = -150\,\text{V}$.

In addition to tuning the switching field $H_c$, $V_g$ also changes the TMR ratio, as shown in Fig. 2(c)-(d). It increases from 90% at $V_g = 0$ to 95% at $V_g = +150\,\text{V}$. We believe there are two main factors that contribute to it. One is that the strain makes the magnetizations in the soft layer and hard layer align better along the easy axis ($y$-axis) due to the enhancement of the magnetic anisotropy when the gate voltage is positive. The other is the modification of the MgO tunnel barrier by the strain since the quantum transport properties of the MTJ could be significantly changed by even a small stretching/squeezing of the crystalline lattice of MgO.[42]

Next, the voltage controlling is investigated in Configuration II, where the MTJ is flanked by a pair of side gates (Fig. 3(a)). In this scheme, the E-field is generated directly underneath the two side gates. The simulated strain mapping is presented in Fig. 3(b). As one can see, when a positive $V_g = +50\,\text{V}$ is applied, the strain fields are formed due to the out-of-plane expansion and in-plane contraction of the region underneath the side gates. In the central gap between the pair of side gates, a strong anisotropic strain ($\varepsilon_{xx} - \varepsilon_{yy} > 0$) is produced with a tensile component $\varepsilon_{xx}$ and a compressive component $\varepsilon_{yy}$, resulting from the interaction of the strain fields under the side gates.[36] In this case, the sign of $\varepsilon_{xx} - \varepsilon_{yy}$ exerted on the MTJ is opposite to that of Configuration I. Hence, the modification of the behavior of MTJ by the gate voltage (Fig. 3(c)-(d)) is opposite to that of Configuration I (Fig. 2(c)-(d)), as expected. A gate voltage of $V_g < 0$ results in $\varepsilon_{xx} - \varepsilon_{yy} < 0$; therefore the magnetic anisotropy of CoFeB layers is enhanced along the $y$-axis and the switching field is increased by ~4-fold from 25 Oe ($V_g = 0$) to 95 Oe ($V_g = -150\,\text{V}$). Similarly, $V_g > 0$ leads to $\varepsilon_{xx} - \varepsilon_{yy} > 0$ and consequently the MR loop becomes slanted, due to the induced magnetic anisotropy along the $x$-axis. Moreover, the TMR ratio slightly increases upon application of a negative gate voltage as shown in Fig. 3(d).

Finally, we have demonstrated strain-induced MTJ toggling by applying gate voltage pulses of ±80 V (Fig. 4(d)). This experiment is performed with Configuration II, and the variation of MR loop from $V_g = -80$ V to $V_g = +80$ V (as shown in Fig. 4(c)) is consistent with the result in Fig. 3(c). A micromagnetic simulation has been performed (Fig. 4(a)-(b)) utilizing the Object Oriented Micro Magnetic Framework (OOMMF)[43] to help understand the MTJ toggling. At $V_g = -80$ V (Fig. 4(a)), the magnetizations of both hard and soft CoFeB layers in the MTJ become parallel along the $y$-axis (with a small bias field $H = 30$ Oe applied along +$y$-direction to overcome any dipole interaction), leading to the low-resistance state denoted by the blue arrow in Fig. 4(c). Once $V_g$ changes to +80 V, magnetizations of the soft layer and the hard layer rotate towards the ± $x$-directions (i.e. opposite directions) because of the generated strain (which overcomes both shape anisotropy and the bias magnetic field). They rotate in opposite directions because of the magnetostatic dipole coupling between the layers, which favors their anti-parallel alignment (Fig. 4(b)). This increase in the angular separation between the magnetizations of the two layers results in a high-resistance state for the MTJ. When the voltage is switched back to -80V, the magnetizations of the two layers again become parallel along the +$y$-direction because of the bias magnetic field, and the MTJ resistance drops. Therefore, by alternative application of the gate voltages of +80 V and -80 V, the MTJ cell can be toggled between high (anti-parallel) and low (parallel) resistance states.

It should be noticed that, although both the soft and hard layer gets altered with $V_g = +80$ V in Fig. 4, the rotation of the soft layer is easier because of its smaller shape anisotropy. The soft layer is thinner and thus has larger out-of-plane demagnetization factor $N_d^z$. Hence, the shape anisotropy field (which is proportional to the difference in the two in-plane demagnetization factors, $N_d^x - N_d^y$) is smaller in the soft layer than for the hard layer. Consequently, if the two layers see comparable levels of stress (in other word the same stress anisotropy field), the stress anisotropy is able to beat the shape anisotropy more effectively in the soft layer than in the hard layer.

There are several advantages of using our local gating scheme compared to previous approaches.[34,40] A tunable anisotropic in-plane strain can be generated in an isotropic piezoelectric material with our local gating scheme. Hence, having a piezoelectric single crystalline substrate with a specific orientation (like PMN-PT (011)) is no longer a necessity to provide an anisotropic strain profile. Moreover, having a highly localized strain in the MTJ region helps to overcome the substrate clamping issue for next generation strain-MRAM devices using a piezoelectric thin film (like PZT) grown on a Si substrate.[35,36] Additionally, it has been predicted that with one more pair of side gates, a deterministic 180° reversal of the magnetization can be achieved.[19,32] The side-gated MTJ demonstrated in our work paves the way towards this fully strain-induced MTJ switching in a double gating device.

One of the key advantages of our proposed device is the reduction of the operating voltage. The typical value of E-field required for 90° rotation of magnetization is ~8 kV/cm, [23,25,26,34] corresponding to a 400 V voltage applied across the 0.5mm-thick substrate. In our experiments, however, the switching voltage $V_g$ reduces to around 100 V, owing to the concentration of the E-field in the local gating scheme (see supplementary material[39]). Had we used a piezoelectric thin film of ~100 nm thickness deposited on a Si substrate as opposed to a 0.5 mm thick substrate used here, the gate voltages $V_g$ would have been reduced by a factor of roughly 5,000 to about 20 mV. Although there are no reports of switching the magnetization of nanomagnets on a 100 nm piezoelectric film, there is a recent report of controlling the states of nanomagnets on a 1000 nm thick piezoelectric film deposited on Si substrate.[30] In such a clamped thin film, the piezoelectric coefficient dropped by 40%. If we assume an 80% drop in the piezoelectric coefficient in a 100 nm thin film, then the gate voltage will increase five-fold to 100 mV. The gate capacitance $C$ has been estimated in previous works to be about 2 fF depending on the dimensions of the electrodes.[19,32] Hence the energy dissipated to toggle the MTJ resistance would have been $CV_g^2 = 20$ aJ, which would make our proposal the lowest energy writing scheme existent.

In summary, we have demonstrated a giant voltage manipulation of CoFeB/MgO/CoFeB MTJ deposited on PMN-PT (001) substrate by using local gating scheme for strain generation. The generated strain is anisotropic and highly localized in the MTJ region which is also confirmed by simulation results. Application of tensile strain along the easy axis (and/or compressive strain along the hard axis) increases the magnetic anisotropy resulting in a significant increase (by a factor of 4) in the switching field of the MTJ. Application of strains of the opposite sign decreases, and ultimately overcomes, the magnetic shape anisotropy, causing a 90° rotation of the magnetization away from the easy axis. Thus, by applying a voltage of alternating sign, which generates strains of alternating signs, we were able to toggle an MTJ between high/low-resistance states. The demonstration of highly effective voltage manipulation of MTJ via localized strains paves the way towards deterministic 180° MTJ switching, and represents a key step towards realizing realistic strain-based MRAM with write energy of few tens of aJ/bit provided the piezoelectric properties scale to 100 nm thickness.

## ACKNOWLEDGMENTS

This work was partially supported by the Center for Spintronic Materials, Interfaces and Novel Architectures (C-SPIN), one of six SRC STARnet Centers. N. D'Souza and J. Atulasimha were partially supported by NSF CAREER grant CCF-1253370. S. Bandyopadhyay was supported by the NSF grant ECCS 1124714.

**Figure Captions**

FIG. 1. (a) Schematic of the strained-MTJ device. (b) An optical micrograph image of the actual fabricated device. (c) In-plane strains in a bare PMN-PT (001) substrate as a function of applied electric field (voltage). The vertical jumps at the maximum E-fields result from the settling of the sample at those points for 10 min. (d) Magnetic hysteresis loop of patterned MTJ films and the MR loop of a MTJ device on PMN-PT without gate voltage application. The magnetic field is along the major axis of the pillars in the *y*-direction.

FIG. 2. Results for strained-MTJ in Configuration I. (a) Schematic of Configuration I. (b) Simulation result showing the mapping of the in-plane anisotropic strain $\varepsilon_{xx} - \varepsilon_{yy}$ upon application of the gate voltage of $V_g = +50\,\text{V}$. The solid line ellipse at the origin denotes the MTJ pillar, and the dashed lines denote the positions of electrodes and side gates. (c) Experimental magnetoresistance (MR) curves characterized under different gate voltages. (d) Measured variation of the switching field (square-line) and TMR ratio (circle-line) of the MTJ as a function of $V_g$.

FIG. 3. Results for strained-MTJ in Configuration II. (a)-(d) are similar to those in Fig. 2.

FIG. 4. Demonstration of voltage manipulation of MTJ toggling (in Configuration II). (a)-(b) Micromagnetic simulation results demonstrating the magnetization configuration of hard and soft CoFeB layers after application of (a) $V_g = -80\,\text{V}$ and (b) $V_g = +80\,\text{V}$. The dimension of the magnet is $3\,\mu\text{m} \times 6\,\mu\text{m}$. Black arrows indicate the direction of magnetic moments. (c) MR loops for $V_g = -80\,\text{V}$ and $V_g = +80\,\text{V}$. The blue arrow indicates the switchable high- and low-resistance states. (d) Toggling of the MTJ between high- and low-resistance states with application of ±80 V gate voltage pulsing. A small bias magnetic field of 30 Oe is applied along the +$y$-axis to overcome the dipole interaction between the two magnetic layers.

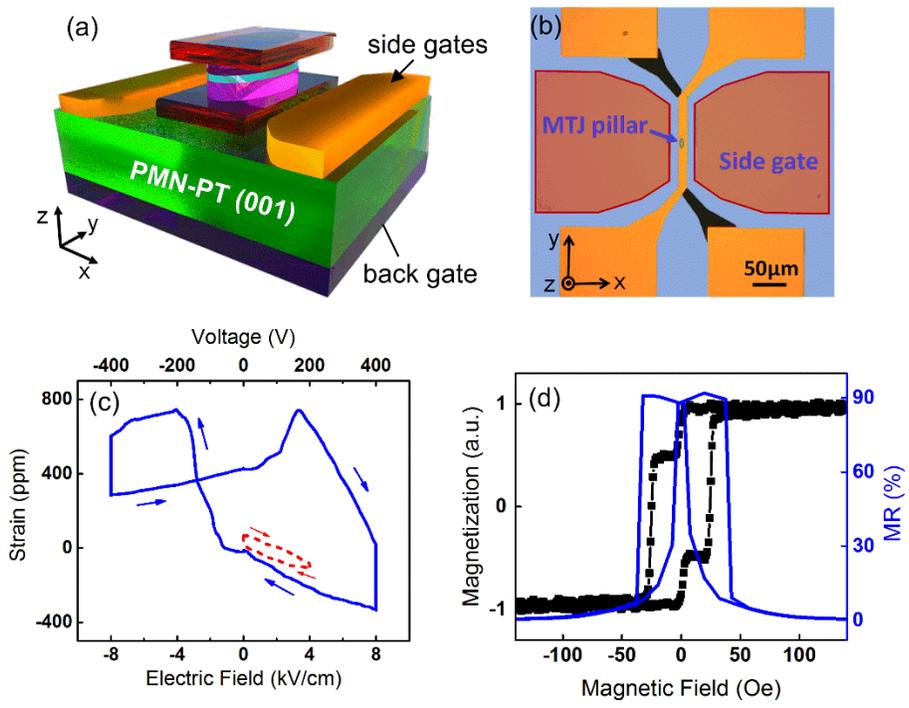

**Figure 1.**

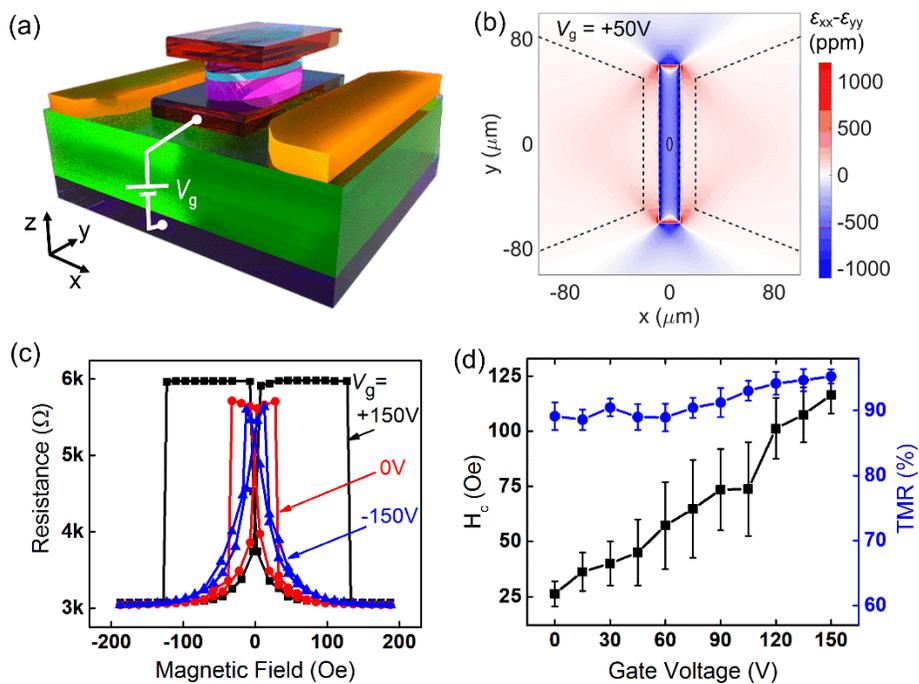

**Figure 2.**

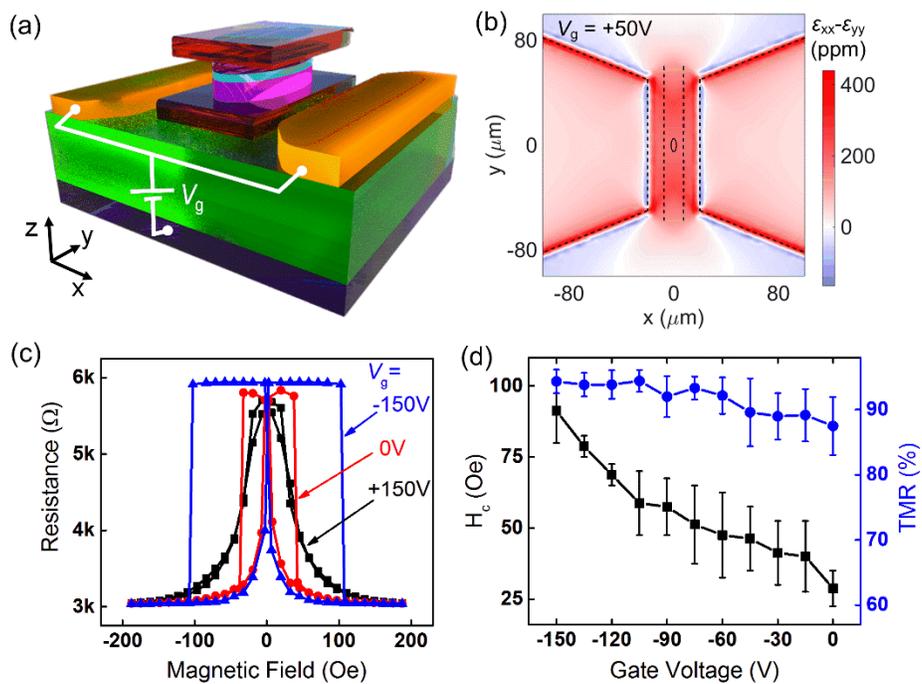

**Figure 3.**

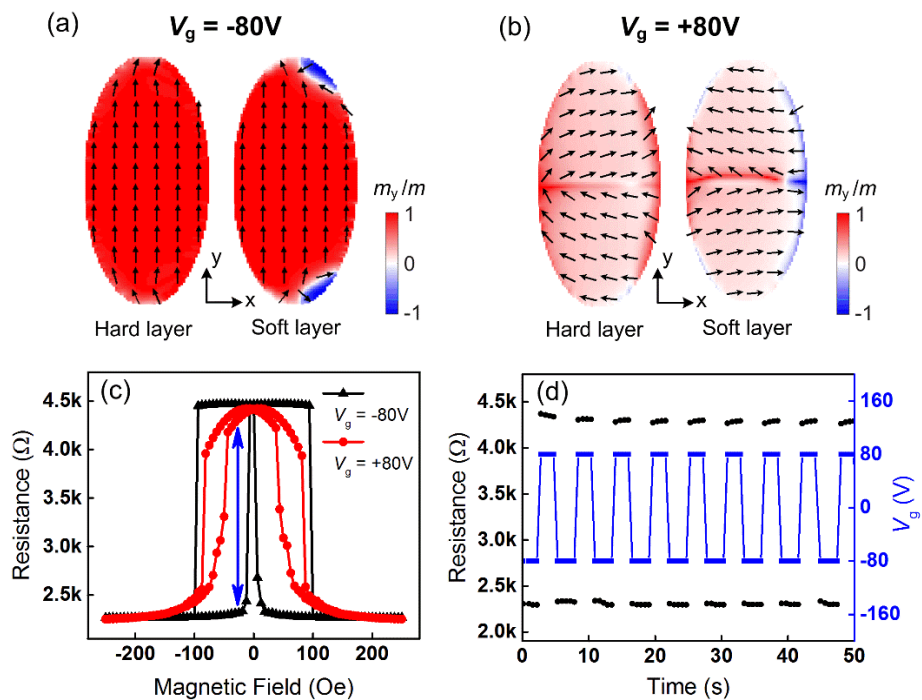

**Figure 4.**

Supplementary Material

# Giant Voltage Manipulation of MgO-Based Magnetic Tunnel Junctions via Localized Anisotropic Strain: a Potential Pathway to Ultra-Energy-Efficient Memory Technology

*Zhengyang Zhao[1], Mahdi Jamali[1], Noel D'Souza[2], Delin Zhang[1], Supriyo Bandyopadhyay[3], Jayasimha Atulasimha[2], and Jian-Ping Wang[1]\**

[1]Department of Electrical and Computer Engineering, University of Minnesota, MN 55455

[2]Department of Mechanical and Nuclear Engineering, Virginia Commonwealth University, Richmond, VA 23284

[3]Department of Electrical and Computer Engineering, Virginia Commonwealth University, Richmond, VA 23284

*Electronic address: jpwang@umn.edu

**A. Sample Preparation and Experimental Setup**

**B. Piezoelectric Finite Element Simulations in the Local Gating Scheme**

    *1. Simulation results for the electric field*

    *2. Simulations to examine the different strain components*

    *3. Discussion about the high efficiency of strain generation in the local gating scheme*

**C. Additional Experimental Results of Voltage-Controlled Modification in MR Loop**

    *1. Voltage-controlled modification in major MR loop*

    *2. Voltage-controlled modification in minor MR loop*

    *3. Results for a second sample with the positions of soft layer and hard layer interchanged*

**D. Calculation of Strain-induced Magnetic Anisotropy**

**E. Discussion about the Scalability**

## A. Sample Preparation and Experimental Setup

The MTJ films, with the structure (from bottom to top, thicknesses in nm) Ta(8)/CoFeB(10)/MgO(1.8)/CoFeB(4)/Ta(8), were directly deposited on the PMN-PT(001) substrate by ultra-high vacuum DC and RF magnetron sputtering. Before the film deposition, the PMN-PT substrate was electrically polarized along the out-of-plane direction with an electric field of 8 kV/cm. The MTJ devices were fabricated using photolithography and Ar ion milling. The area of the MTJ pillars ranges from 16 to 40 $\mu m^2$. A post-annealing process was performed in vacuum under a magnetic field of about 0.4 T at 250 °C for 1 hour, to improve the crystallization of the MgO tunnel barrier.

The resistance of MTJ was measured at room temperature using the four-probe technique under a bias current of 5 $\mu A$. A Keithley 6221 current source generated the dc-current and the output voltage was characterized using a Keithley 2182 nano-voltmeter. The gate voltage was applied with a Keithley 2400 sourcemeter. The MR loops were obtained by sweeping the magnetic field along the $y$-direction (long-axis of the MTJ).

## B. Piezoelectric Finite Element Simulations in the Local Gating Scheme

In the main text we present the strain mapping for Configuration I (Fig. 2(b)) and II (Fig. 3(b)), obtained via the piezoelectric finite element simulations. In this section, more details of the simulations are presented, including the electric field distribution in the piezoelectric layer, and the mapping of the three strain components $\varepsilon_{xx}$, $\varepsilon_{yy}$ and $\varepsilon_{zz}$, respectively. Based on the simulation results, it will be shown that local gating scheme can be scaled down to nano-scale.

The finite elements model was developed using COMSOL Multiphysics to approximate the E-fields and strains observed around the MTJ device. The PMN-30PT was modeled as a 1.5×1.5×0.5mm element using the piezoelectric module. To decrease the complexity of the simulation, all the deposited thin films, including the MTJ and the electrode layer, were not considered in the simulation (this will not change the simulation results too much because the thicknesses of the MTJ films are negligible compared with the PMN-PT substrate). The PMN-30PT material properties are:[1] $d_{33} = 1981 \times 10^{-12}$ C/N, $d_{31} = -921 \times 10^{-12}$ C/N, and $\rho = 8.043 \times 10^3$ kg/m$^3$. All the boundaries of the PMN-PT element are mechanically free except the bottom surface of the element which is restricted in the z=0 plane. A voltage of $V_g = +50$ V is applied to the top contacts with the bottom surface grounded (according to gating scenarios in Configuration I/II); that is, the electric field is applied through the thickness of the PMN-PT.

*B1. Simulation results for the electric field*

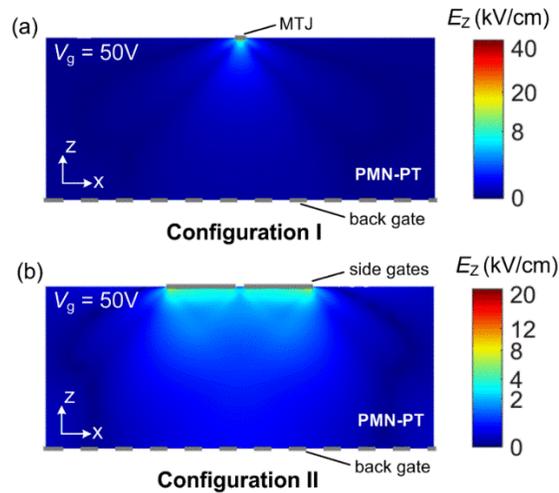

**Fig. S1.** Simulation demonstrating the out-of-plane electric field strength within the $y = 0$ cross-section, with the application of $V_g = +50$ V . a) Configuration I, b) Configuration II. The thickness of PMN-PT is 0.5mm.

The distribution of the electric field when applying the gate voltage in Configuration I/ II is simulated and presented in Fig. S1. Due to the small dimensions of the electrodes and side gates compared with the large common back gate, the E-field is highly concentrated. In Configuration I, the E-field is concentrated under the MTJ bottom electrode; in Configuration II, the E-field is concentrated just beneath the side gates. The highly concentrated E-field is the reason for the highly localized strain profile generated in the substrate. The concentration of the E-field also allows us to use a relative small voltage to generate a large strain for MTJ controlling (see Table S1).

*B2. Simulations to examine the different strain components*

In Figs. 2(b) and 3(b) of the main text, we present the mapping of the anisotropic strain, i.e. $\varepsilon_{xx} - \varepsilon_{yy}$, on the top surface of the sample. In this section, we will show the distribution of each strain component. Figure S2 gives the simulation results showing the mapping of the strains along $x$, $y$ and $z$ directions, with the same $V_g = +50\,\text{V}$ applied. For Configuration I (Fig. S2(a)-(c)), component $\varepsilon_{xx}$ is dominant over $\varepsilon_{yy}$ in the MTJ region, since the electrode is a narrow stripe along the $y$-direction. Therefore, the strain exerted on the MTJ is anisotropic. By subtracting the value of $\varepsilon_{yy}$ at each spatial point from $\varepsilon_{xx}$ at the corresponding point, we get the distribution of $\varepsilon_{xx} - \varepsilon_{yy}$ shown in Fig. 2(b) in the main text.

For Configuration II (Fig. S2(d)-(f)), the interaction of the strains underneath the pair of side gates induces an anisotropic strain field in the central gap,[2] where the strain along $x$-axis is tensile ($\varepsilon_{xx} > 0$, see Fig. S2(d)) and strain along $y$-axis is compressive ($\varepsilon_{yy} < 0$, see Fig. S2(e)) for a positive gate voltage. The distribution of $\varepsilon_{xx} - \varepsilon_{yy}$ is given in Fig. 3(b) in the main text.

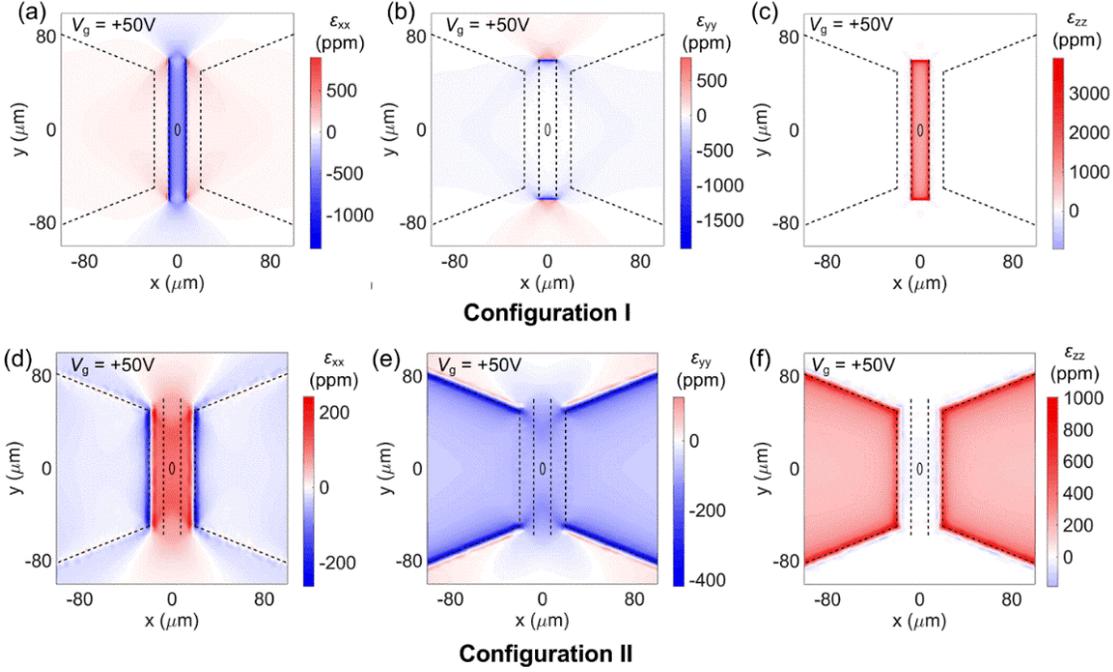

**Fig. S2.** Finite element simulation showing the mapping of different piezoelectric strain components on the top surface of the substrate, with the application of $V_g = +50\,\text{V}$. (a),(d) in-plane strain component $\varepsilon_{xx}$, (b),(e) in-plane strain component $\varepsilon_{yy}$, (c),(f) out-of-plane strain component $\varepsilon_{zz}$. (a)-(c) are for Configuration I and (d)-(f) are for Configuration II.

### B3. Discussion about the high efficiency of strain generation in the local gating scheme

Table S1 summarizes the strains produced in the MTJ region by using different gating schemes. The values of strains are obtained from the simulation results. Clearly, in our local gating schemes (Configurations I and II), the strain generation is much more efficient than in uniform gating scheme (applying voltage on uniform contacts).[3] The localized strain $\varepsilon_{xx} - \varepsilon_{yy}$ exerted on the MTJ is 3.0 times (Configuration I) or 1.4 times (Configuration II) larger than the in-plane strain that can be produced by uniform gating. Therefore, by using the local gating design, not only can we get the required anisotropic strain from isotropic piezoelectric materials, but the strain generation efficiency also improves significantly.

**Table S1**. Strain generated by using different gating schemes (with $V_g = +50\,\text{V}$)

| Gating schemes | In-plane Strain in MTJ region | Out-of-plane Strain in MTJ region |
|---|---|---|
| **Configuration I** | -274 ppm *(anisotropic)* | 618 ppm |
| **Configuration II** | 123 ppm *(anisotropic)* | n/a |
| **Uniform Gating** | -92 ppm *(isotropic)* | 196 ppm |

The high efficiency of strain generation in the local gating scheme can be attributed to two factors. First, for uniform gating, the E-field is uniformly distributed everywhere within the substrate. Whereas by using the local gating scheme, the E-field is highly concentrated in the regions of interest and the field around the MTJ is much stronger (Fig. S1). Second, in Configuration II, the in-plane strains along the $x$-direction and along the $y$-direction have opposite signs and can compensate each other, making the anisotropic strain $\varepsilon_{xx} - \varepsilon_{yy}$ larger. Obviously, by further optimizing the dimensions of the contacts in the local gating scheme, the strain generation efficiency can be further improved.

## C. Additional Experimental Results of Voltage-Controlled Modification in MR Loop

In Fig. 2(c) and 3(c) of the main text, we pick up three data points, at $V_g = +150\,\text{V}$, $0$ and $-150\,\text{V}$, to show the voltage modification in the MR (major) loop of MTJ. In this section, MR loops generated under additional gate voltages are presented to make the variation trends clearer. Apart from the major MR loops, the variations in the minor MR loops are also shown to illustrate the voltage control of the magnetic anisotropy of the soft layer. Moreover, the results for a second sample with the position of the soft layer and hard layer interchanged is presented.

### C1. Voltage-controlled modification in major MR loop

As supplement to Fig. 2(c)-(d) and Fig. 3(c)-(d) of the main text, Fig. S3 presents the MR curves under a series of gate voltages, gradually changed from $+150\,\text{V}$ to $-150\,\text{V}$. Accordingly, the variation of the switching field and TMR ratio is shown for both positive and negative $V_g$. In Fig. S3(a) (Configuration I), the switching field of the MTJ gradually increases with positive $V_g$ indicating the enhancement of the magnetic anisotropy along the easy axis, and the MR loop becomes slanted with a negative $V_g$ indicating the change of the easy axis direction. In the negative $V_g$ region, the sharp switching disappears, so the "switching field" is determined by the point with the largest $\left| dR/dH \right|$ value in the MR loops. Although such "switching field" is almost constant in this region, the saturation field keeps increasing with the negative $V_g$ (Fig. S3(a)), indicating the enhancement of magnetic anisotropy along the hard axis of the MTJ. The MR ratio stays at a low value in the negative $V_g$ region, for the reasons we have discussed in the main text. For Configuration II (Fig. S3(c)-(d)), the trends are opposite.

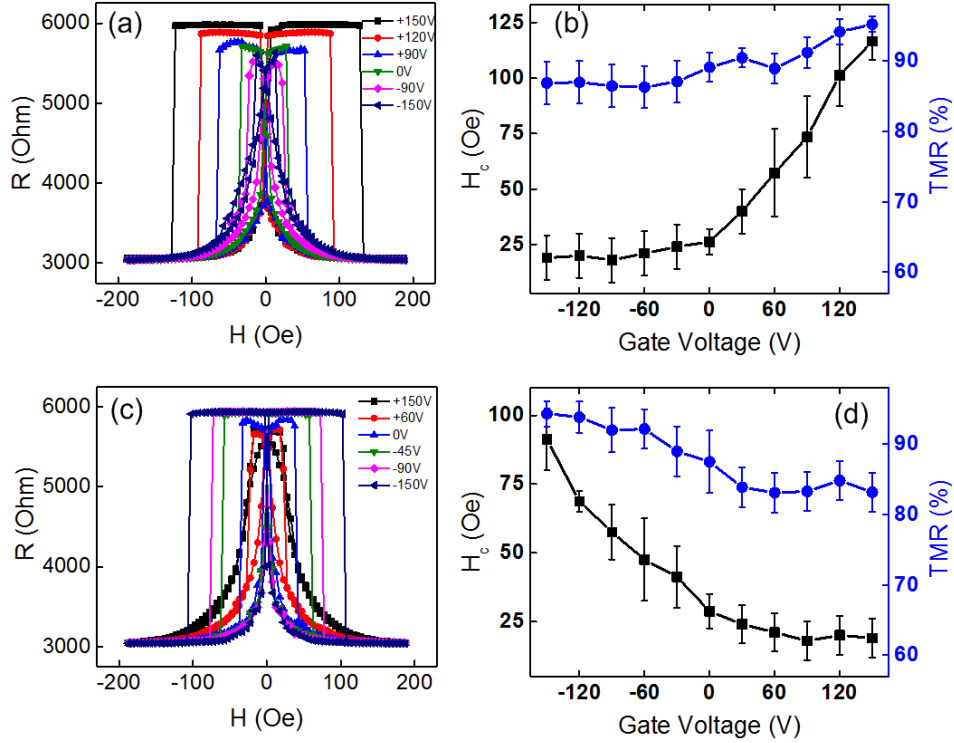

**Fig. S3.** Major MR loops characterized under different gate voltages in a) Configuration I and c) Configuration II, and variation of the switching field (square-line) and TMR ratio (circle-line) of the MTJ as a function of $V_g$ changing from -150V to +150V.

### C2. Voltage-controlled modification in minor MR loop

The strain modification in the soft layer of MTJ is not observable from the MR major loops in Fig. 2(c) and 3(c) in the main text. It is because the dipolar field from the hard layer plays a dominant role in the switching of the soft layer. In order to illustrate that both CoFeB layers of the MTJ can be controlled by voltage, we measured the MR minor loop with different $V_g$ applied, as shown in Fig. S4. The MR minor loops were obtained by performing a sweep of the magnetic field in a small range (from -100 Oe to 15 Oe), during which only the soft CoFeB layer can be flipped by the magnetic field.

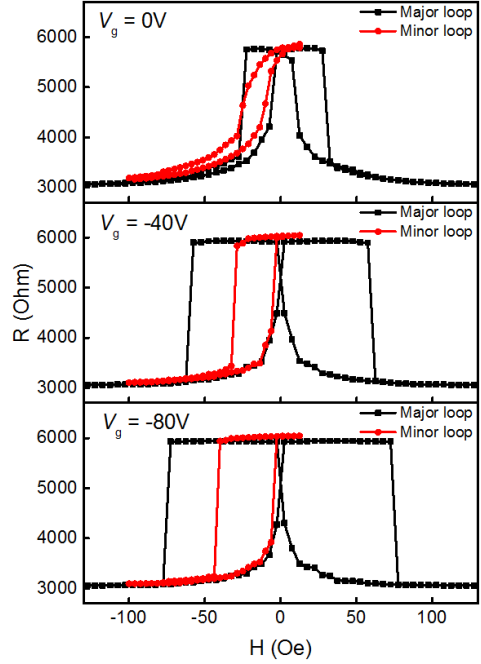

**Fig. S4.** MR major and minor loops with (a) $V_g = 0\,\text{V}$, (b) $V_g = -40\,\text{V}$, and (c) $V_g = -80\,\text{V}$ (in Configuration II).

In Fig. S4, as $V_g$ is changed from 0 to +80V, not only does the switching field of the hard CoFeB layer increase (from the major loops), but the coercivity of the soft layer also increases from 7 Oe to 18 Oe (from the minor loops). Therefore, we can conclude that the magnetic anisotropy of both CoFeB layers in the MTJ can be effectively controlled by the gate voltage.

*C3. Results for a second sample with the positions of soft layer and hard layer interchanged*

In the main text, we have mentioned that the magnetic anisotropy of both CoFeB layers can be affected by the strain, because neither layer is pinned by an anti-ferromagnetic layer. And we argue that the strain exerted on the soft layer and that exerted on the hard layer is very close to each other, although the hard layer (CoFeB 10nm) is closer to the piezoelectric substrate. This is because the strain relaxation between layers can be negligible, considering the thickness of the MTJ stack is much smaller than its lateral dimensions.

To confirm this, we have made a second sample with the structure of Ta(8)/CoFeB(4)/MgO(1.8)/CoFeB(10)/Ta(8) (from bottom to top, thickness in nm), so that the soft layer now is closer to the substrate. The results of this sample is presented in Fig. S5. It can be seen the variations of the MR loop are quite similar to the results for the sample Ta(8)/CoFeB(10)/MgO(1.8)/CoFeB(4)/Ta(8). Therefore, for our devices, interchanging the positions of the soft layer and hard layer doesn't make too much change.

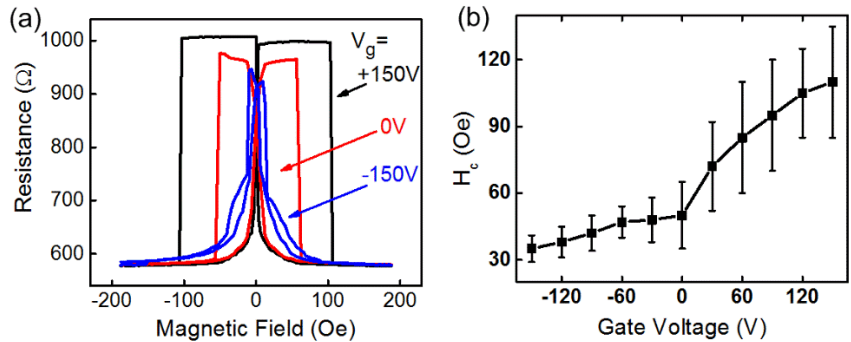

**Fig. S5.** MR loops characterized under different gate voltages (in Configuration I) for the sample Ta(8)/CoFeB(4)/MgO(1.8)/CoFeB(10)/Ta(8), where the soft layer is closer to the piezoelectric substrate.

However, it should be noted that for a further scaled devices (with the MTJ lateral dimensions of sub-100 nm), the strain relaxation needs to be considered, and the soft (free) layer should be deposited closer to the piezoelectric layer in order to experience maximal strain.

## D. Calculation of Strain-induced Magnetic Anisotropy

From the MR loops in Fig. 2(c) and Fig. 3(c) in the main text, the magnetic anisotropy induced by strain can be evaluated. Consider Configuration II, for example. When no gate voltage is applied, the anisotropy of the CoFeB layers comes from the shape anisotropy, and the anisotropy field can be determined by the saturation field of the MR loop with magnetic field swept along the short axis (Fig. S6), which is $H_K^{shape} = 120\,\text{Oe}$. With a positive or negative gate voltage applied, the strain will induce an anisotropy. So the total anisotropy field would be: $H_K = H_K^{shape} + H_K^{strain}$. Assuming the switching field $H_c \propto H_K$, since $H_c$ increases 3.5 times from $V_g = 0$ to $V_g = 150\,\text{V}$ in Fig. 3(d) in the main text, $H_K^{strain} \approx 2.5 \times H_K^{shape} \approx 300\,\text{Oe}$ for $V_g = 150\,\text{V}$. Therefore the strain-induced anisotropy $K_{me} = \frac{1}{2} H_K^{strain} M_s = 150 \times 10^3\,\text{erg/cm}^3$ for $V_g = 150\,\text{V}$ in Configuration II. Note that this is just a rough calculation, since the saturation field obtained from Fig. S6 may not equal the shape anisotropy field of the hard layer (because of the interaction of the two layers), and the assumption of $H_c \propto H_K$ is debatable for the multi-domain MTJ. But the order of magnitude of $K_{me}$ calculated in this section is reliable.

Finally, using this calculated $K_{me}$, we can further estimate the value of the strain produced on the MTJ. Considering that $K_{me} = \frac{3}{2}\lambda\sigma = \frac{3}{2}\lambda Y(\varepsilon_{xx} - \varepsilon_{yy}) = 150 \times 10^3\,\text{erg/cm}^3$, $Y = 160\,\text{GPa}$, and $\lambda = 3 \times 10^{-5}$ for CoFeB thin film,[8,9] the anisotropic strain applied to the MTJ is determined to be about $\varepsilon_{xx} - \varepsilon_{yy} = 1950\,\text{ppm}$ at $V_g = 150\,\text{V}$ which is consistent with the simulation result in Fig. 3(d).

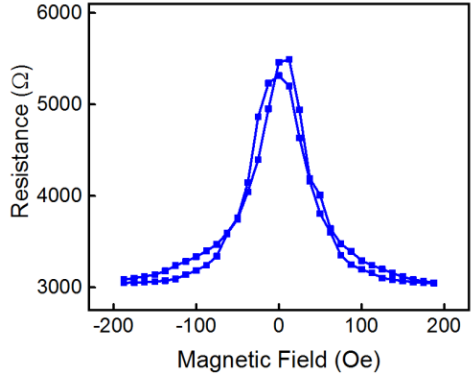

**Fig. S6.** MR curve without application of gate voltage and magnetic field swept along the minor axis.

## E. Discussion about the scalability

As we mentioned in the main text, the realistic strain-mediated MRAM requires using a piezoelectric thin film deposited on a traditional Si substrate to generate the strain.[2,4,5] The device should be scaled down to sub-100 nm.[6] In this section we examine the scalability of the side-gated MTJ device. In our experiments, the PMN-PT layer (substrate) is 0.5 mm in thickness. If it is scaled down to a piezoelectric thin film of 100 nm in thickness (reduced by a factor of 5000), the lateral dimensions of the side gates can be correspondingly scaled from ~200 µm to 50 nm (in this case, the lateral dimensions of the side gates are in the same level as the dimension of the MTJ pillar, and the strain generation is still efficient[5]). If we consider an 80% drop in the piezoelectric coefficient in a 100 nm thin film due to the substrate clamping effect,[7] then the gate voltage to switch the MTJ resistance (and write bits)  will be on the order of 100 mV, as mentioned in the main text. This will result in extremely low write energy of few tens of aJ/bit, leading to an ultra-energy-efficient non-volatile memory technology.